\documentclass[
 numbers ]
  {aipproc}

\layoutstyle{6x9}
\usepackage{amsmath}

\begin{document}

\title{Effective Conformal Descriptions\\of Black Hole Entropy: A Review}


\author{Steven Carlip}{
  address={Department of Physics, University of California at Davis, Davis, CA 95616, USA}
}

\begin{abstract}
Black holes behave as thermodynamic objects, and it is natural to ask for an
underlying ``statistical mechanical'' explanation in terms of microscopic 
degrees of freedom.  I summarize attempts to describe these degrees of freedom
in terms of a dual two-dimensional conformal field theory, emphasizing the 
generality of the Cardy formula and the consequent universal nature of the 
conformal description.
\end{abstract}

\maketitle


\section{Introduction}

Black holes are not black.  Rather, they are black bodies: as Bekenstein \citep{Bekenstein} 
and Hawking \citep{Hawking} showed some 40 years ago, they ``glow'' with a
characteristic temperature and entropy
\begin{equation}
T_H = \frac{\hbar\kappa_H}{2\pi}, \qquad
S_{\scriptscriptstyle\mathit{BH}} = \frac{A}{4\hbar G},
\label{BH}
\end{equation}
where $\kappa_H$ is the surface gravity ($1/4GM$ for the Schwarzschild case) and
$A$ is the area of the event horizon.  While this thermodynamic behavior 
has not been directly observed---the Hawking temperature for an astrophysical black 
hole is far too small---it can be derived in so many different ways \citep{Carlip} that 
there is very little doubt of its validity.

In normal thermodynamic systems, temperature and entropy are collective 
properties of microscopic states.  It is natural to expect the same to be true 
for black holes.  If it is, an understanding of black hole thermodynamics 
could offer precious insights into quantum gravity.  Indeed, the Bekenstein-Hawking 
entropy formula is one of the very few relationships we know that is genuinely 
quantum gravitational, in the sense that it contains both Planck's constant 
$\hbar$ and Newton's constant $G$.

In this article, I will briefly summarize the ``black hole CFT'' program, an attempt 
to describe the microphysics of black holes in terms of a ``dual'' two-dimensional 
conformal field theory.  For  more details, I refer the reader to the review article 
\citep{Carlip2}.

\section{The puzzle of universality}
 
We start with one of the outstanding mysteries of black hole thermodynamics, 
the problem of ``universality.''  This may  be seen as a pair of puzzles:
why do so many different microscopic descriptions of black hole entropy give 
the same answer, and why does this answer take such a simple and general 
form?

First: many very different models of microscopic degrees of freedom give the 
same black hole entropy.  These include
\begin{itemize}
\item three approaches in string theory: weakly coupled D-branes \citep{Strominger},
the AdS/CFT correspondence \citep{AGMOO}, and the ``fuzzball'' picture \citep{Mathur};
\item three approaches in loop quantum gravity: boundary Chern-Simons theory 
\citep{ABK}, counting intertwiners \citep{Livine}, and a ``gas of punctures'' \citep{Ghosh};
\item counting the ``heavy'' states of induced gravity \citep{Frolov};
\item holographic entanglement entropy \citep{Ryu};
\item computations from a single instanton \citep{Gibbons} and from pair production
\citep{Garfinkle};
\end{itemize}
and a number of others \citep{Carlip}.  None of these models is complete---we do
not, after all, have a finished quantum theory of gravity---and all involve added
assumptions or restrictions.  But within its realm of validity, each approach 
reproduces the standard Bekenstein-Hawking entropy.   In some cases, we may 
simply have different depictions of the same microscopic degrees of freedom; 
the string theory descriptions, for instance, are presumed to be ``dual'' to 
each other.  But others seem quite different, and their universal agreement is 
puzzling.

Second: the final form of the Bekenstein-Hawking entropy is also surprisingly simple
and universal.  The area law holds for black holes in any dimension, with any set of
charges and spins.  The same relationship holds for the more elaborate ``black''
objects that can exist in higher dimensions: black strings, black rings, black saturns
and the like \citep{Emparan}.  One can change the entropy by changing the action,
but even then the new expression takes a simple form in terms of the revised action
\citep{Wald}.

Together, these two features strongly suggest that there should be a
simple underlying structure that controls the statistical mechanics of the quantum
black hole, one that is general enough to determine the thermodynamic
properties independent of the details of the degrees of freedom.  In the next section, 
I will discuss a rather strange possibility: that this underlying structure is the
conformal symmetry of a two-dimensional theory that is in some sense dual to the
standard description of the black hole.

\section{Two-dimensional conformal field theory}

The idea of describing black hole microstates in terms of a two-dimensional conformal
field theory was first suggested in \citep{Carlip3}, but more as a hope than a concrete
proposal.  The first realization of the idea was achieved by Strominger \citep{Strominger2}
and, simultaneously, by Birmingham, Sachs, and Sen \citep{BSS} for the (2+1)-dimensional
BTZ black hole.  These results were quickly extended to a variety of near-extremal
black holes with near-horizon geometries that look like that of the BTZ solution
(for a review, see \citep{Peet}), and several authors suggested methods for extending
the results to nonextremal black holes \citep{Solodukhin,Carlip4,Carlip5}.  Interest in
the program was revived in the past few years with the discovery of an extremal Kerr/CFT
correspondence \citep{Guica}, and a new, relatively ``clean'' approach to the nonextremal
Kerr black hole was found in 2011 \citep{Carlip6}.

Such an idea raises two obvious questions: why conformal, and why two dimensional?  
We have at least partial answers to both.
\begin{itemize}
\item The near-horizon region of a black hole really \emph{is} approximately conformally
invariant: to an observer who remains outside the horizon, dimensionful quantities
such as masses are rapidly red-shifted away \citep{Camblong}, and the near-horizon
region of a generic black hole admits an approximate conformal Killing vector \citep{Martin}.
\item The same red shift that eliminates masses also makes transverse excitations negligible 
relative to those in the $r-t$ plane, leaving two ``important'' dimensions.  This fact 
has been put to good use in calculations of Hawking radiation, which can be carried 
out entirely in the framework of a two-dimensional near-horizon conformal field 
theory \citep{Iso,Bonora}.
\end{itemize}
To a certain extent, though, the existence of a two-dimensional conformal description is 
more a hope than an established result.  Two-dimensional conformal symmetry is extremely 
powerful---it is the only known symmetry strong enough to determine the asymptotic 
density of states of a theory, and thus the entropy.  If the universality of black hole 
entropy is to have a simple explanation, this is a natural place to look.

Any metric on a two-dimensional manifold can be written locally as
\begin{equation}
ds^2 = 2g_{z{\bar z}}dzd{\bar z}
\end{equation}
in terms of complex coordinates $z$ and $\bar z$.
Holomorphic and antiholomorphic coordinate changes $z\rightarrow z + \xi(z)$,
${\bar z}\rightarrow {\bar z} + {\bar\xi}({\bar z})$ merely rescale the metric, and  
provide the basic symmetries of a conformal field theory.   This group of symmetries is
infinite dimensional, distinguishing it from the conformal group of higher dimensional 
spacetimes.  Its generators, denoted $L[\xi]$ and ${\bar L}[{\bar\xi}]$, 
satisfy a Virasoro algebra \citep{CFT},
\begin{align}
&[L[\xi],L[\eta]] 
   = L[\eta\xi' - \xi\eta']   + \frac{c}{48\pi}\int dz\left( \eta'\xi'' - \xi'\eta''\right) \nonumber\\
&[{\bar L}[{\bar\xi}],{\bar L}[{\bar\eta}]] 
   = {\bar L}[{\bar\eta}{\bar\xi}' - {\bar\xi}{\bar\eta}']  + \frac{{\bar c}}{48\pi}%
   \int d{\bar z}\left( {\bar\eta}'{\bar\xi}'' - {\bar\xi}'{\bar\eta}''\right) \nonumber\\
&[ L[\xi],{\bar L}[{\bar\eta}]] = 0 , \label{Vir}
\end{align}
uniquely determined by the values of the two constants $c$ and $\bar c$, the 
central charges.  As in ordinary field theory, the zero modes $L_0$ and ${\bar L}_0$ 
of the symmetry generators are conserved quantities, the ``conformal weights,''
which can be seen as linear combinations of mass and angular momentum.

Two-dimensional conformal symmetry is remarkably powerful.  In particular,
with a few mild restrictions, Cardy has shown that the asymptotic density of
states of any two-dimensional CFT at conformal weight $\{L_0,{\bar L}_0\}$ is 
given by \citep{Cardy,Cardy2}
\begin{equation}
\ln\rho(L_0) \sim 2\pi\sqrt{\frac{cL_0}{6}}, \qquad 
\ln{\bar\rho}({\bar L}_0) \sim 2\pi\sqrt{\frac{{\bar c}{\bar L}_0}{6}} .
\label{micro}
\end{equation}
This result can be viewed as a microcanonical expression for the entropy.  Similarly, 
for a conformal field theory at fixed temperature $T$,
\begin{equation}
\ln\rho(T) \sim \frac{\pi^2}{3}cT, \qquad 
\ln{\bar\rho}(T) \sim \frac{\pi^2}{3}{\bar c}T,
\label{canon}
\end{equation}
a canonical version of the Cardy formula.  The entropy is thus determined by a few 
parameters, regardless of the details of the conformal field theory---exactly the 
kind of universal behavior we would like for black holes.

\section{Building a black hole/CFT correspondence}

General relativity is not, of course, a two-dimensional conformal field theory.  
For the results of the preceding section to be relevant, we must find a Virasoro 
algebra (\ref{Vir}) hidden inside its constraint algebra.  This is most easily 
analyzed in the ADM formalism \citep{Poisson}.  We write the metric in the form
\begin{equation}
ds^2 = -N^2dt^2   
      + q_{ij}\left(dx^i + N^i dt\right)\left(dx^j + N^j dt\right) ,
\end{equation}
with canonical variables $\{q_{ij},\pi^{ij}\}$ and an action  
\begin{equation}
I = \frac{1}{16\pi G}\int dt \int d^{n-1}x\,
     \left[ \pi^{ij}{\dot q}_{ij} - N^i{\cal H}_i - N{\cal H}\right] .
\end{equation}
The constraints ${\cal H}_i$ and $\cal H$ generate diffeomorphisms (or, more properly,
``surface deformations,'' equivalent to diffeomorphisms on shell \citep{Teitelboim})---that 
is, the Hamiltonian $H[\xi^\perp,{\hat\xi}^i] = \int d^3x\left[\xi^\perp{\cal H} 
+ {\hat\xi}^i{\cal H}_i\right]$ has Poisson brackets
\begin{equation}
\left\{ H[\xi],F[q,\pi]\right\} = \delta_\xi F[q,\pi] 
\end{equation}
with any phase space variables.  

On a spatially closed manifold, the Poisson algebra of the generators 
$H[\xi^\perp,{\hat\xi}^i]$ closes, with no central term.  In the presence of a 
boundary, however, the generators acquire boundary terms, 
$H[\xi^\perp,{\hat\xi}^i] \rightarrow {\bar H}[\xi^\perp,{\hat\xi}^i] %
= H[\xi^\perp,{\hat\xi}^i] + B[\xi^\perp,{\hat\xi}^i]$, and these can modify 
the algebra:
\begin{equation}
\{{\bar H}[\xi], {\bar H}[\eta]\} = {\bar H}[\{\xi,\eta\}] + K[\xi,\eta]  .
\end{equation}
The new term $K[\xi,\eta]$ is always central---that is, it has vanishing brackets with all phase
space variables---and it takes a general form computed in \citep{Carlip6},
\begin{align}
K[\xi,\eta] = -B[\{\xi,\eta\}_{\scriptscriptstyle\mathit{SD}}]  
    &-\frac{1}{8\pi G}\int_{\partial\Sigma}\!\!d^{n-2}x \sqrt{\sigma}\,n^k \Bigl[
    -\frac{1}{2}\frac{1}{\sqrt{q}}
               {\hat\xi}_k\eta^\perp\mathcal{H}\nonumber\\
    & - {\hat\xi}_i\eta^\perp\,{}^{(n-1)}\!R^i{}_k
    + D_i{\hat\xi}_kD^i\eta^\perp - D_i{\hat\xi}^iD_k\eta^\perp \nonumber\\
    &+ \frac{1}{\sqrt{q}}{\hat\eta}_k\pi^{mn}D_m{\hat\xi}_n
       +\frac{1}{2}\frac{1}{\sqrt{q}}\pi_{ik}\{\xi,\eta\}_{\scriptscriptstyle\mathit{SD}}^i
        - (\eta\leftrightarrow\xi)\Bigr]  .
\label{center}
\end{align}
We thus obtain a ``recipe'' for a black hole/CFT correspondence:
\begin{enumerate}
\item Identify an appropriate boundary and the corresponding ``black hole'' 
boundary conditions.
\item Compute the boundary terms $B[\xi^\perp,{\hat\xi}^i]$ for the generators 
of diffeomorphisms.  They should be chosen so that the variation $\delta{\bar H}[\xi]$
is well-defined, with no net boundary contributions; the specific form will 
depend on the choice of boundary conditions.
\item Find the central term $K[\xi,\eta]$ in the resulting algebra.
\item Search for a preferred one- or two-dimensional subalgebra of surface deformations
(technically, a $\hbox{Diff}\,S^1$ or $\hbox{Diff}\,S^1\times\hbox{Diff}\,S^1$ 
algebra).  If such a subalgebra exists, general mathematical results imply that it must be a
Virasoro algebra.
\item Read off the central charges, and use the Cardy formula to count the states.
\end{enumerate}

But while this is a recipe of sorts, it is not exactly a ``cookbook'' recipe.  Key questions
remain:
\begin{itemize}
\item What boundary and boundary conditions should we choose?  Results exist for 
boundaries at infinity in the asymptotically AdS case, boundaries of near-horizon 
regions, and stretched horizons.  Ideally, we might want to choose the horizon itself 
as the boundary, but it is a null surface, a feature that significantly complicates the
constraints.
\item What two-dimensional subgroup do we pick?  For the extremal Kerr-Newman
black hole, for instance, there are two known choices, leading to the so-called $J$ and $Q$ 
pictures \citep{Chen}; each gives the correct entropy, but they cannot be treated together.
For the nonextremal Kerr black hole, on the other hand, it was shown in \citep{Carlip6} that 
the analog of the subgroup used in \citep{Guica} gives only half the entropy.
\item What about 1+1 dimensions?  The black hole in (1+1)-dimensional dilaton gravity 
is a perfectly nice object, with well-defined thermodynamic properties \citep{Grumiller}, 
but the canonical methods I have described here fail---there is not enough 
room at the boundary of a one-dimensional spatial slice for these techniques to work.  
A possible alternative is to introduce boundary conditions as explicit constraints 
in the surface deformation algebra \citep{Carlip7}, but this idea is not yet very fully 
developed.
\item What about the rest of black hole thermodynamics?  Counting states is an important 
first step, but we also need Hawking radiation.  We thus need to understand how to couple our
boundary conformal field theory to ``bulk'' matter.  There has been one interesting step
in this direction for the BTZ black hole \citep{Emparan2}, but the overall problem 
remains very poorly understood.
\end{itemize}

\section{An example: the nonextremal Kerr black hole}

The preceding section was a bit abstract, and it may be helpful 
to look at a specific example.  Consider the generic nonextremal stationary black hole 
in four spacetime dimensions \citep{Carlip6}.  Our starting point is the observation by 
Medved, Martin, and Visser \citep{Medved} that the near-horizon metric of such a 
black hole can always be written in the ADM-like form
\begin{equation}
ds^2 = -N^2dt^2  + d\rho^2 
      + q_{\varphi\varphi}\left(d\varphi + N^\varphi dt\right)^2 
      + q_{zz}dz^2 ,
\end{equation}
where $\rho$ is the proper distance from horizon, and that even without using the
field equations, finite curvature at the horizon requires that
\begin{alignat}{3}
&N =  \kappa_{H}\rho 
       + \frac{1}{3!}\kappa_2(z)\rho^3 + \dots  \qquad\qquad&&
q_{\varphi\varphi} = [q_H]_{\varphi\varphi}(z) 
                    +  \frac{1}{2}[q_2]_{\varphi\varphi}(z)\rho^2 + \dots \nonumber\\
&N^\varphi = -\Omega_{H} 
       - \frac{1}{2}\omega_2(z)\rho^2 + \dots &&
q_{zz} = [q_H]_{zz}(z) + \frac{1}{2}[q_2]_{zz}(z)\rho^2 + \dots 
\label{bcs}
\end{alignat}
We would like to choose a ``stretched horizon'' of small $\rho$ as a boundary, taking 
the limit $\rho\rightarrow0$ at the end.  The obvious choice $\rho = \mathit{const.}$
turns out to be awkward, requiring a complicated angular dependence of
the boundary-preserving diffeomorphisms.  Instead, following \citep{Guica}, let us 
choose a surface of angular velocity $\bar\Omega$---which we will allow to approach
the horizon angular velocity $\Omega_{H}$---and demand that modes of the form 
$\xi(\varphi - {\bar\Omega}\,t)$ be lightlike: 
\begin{equation}
g^{ab}\partial_a\xi\partial_b\xi = 0 \ \Rightarrow \ 
     [q_H]_{\varphi\varphi}({\bar\Omega} - \Omega_{H})^2 = \kappa_{H}{}^2\rho^2 .
\end{equation}
One can equivalently demand that the horizon be a ``stretched Killing horizon,'' 
defined by a Killing vector that is invariant under boundary diffeomorphisms
\citep{Carlip6}.

The boundary conditions (\ref{bcs}) are then preserved by diffeomorphisms of the form
\begin{alignat}{3}
&\delta_\xi N = 0 \quad && \Rightarrow\ \
   {\hat\xi}^\rho = -{\bar\varepsilon}\rho\partial_\varphi\xi^t 
    = -\rho{\bar\partial}_t\xi^t\nonumber\\
& \delta_\xi N^\rho = 0 && \Rightarrow\ \
    \rho\partial_\rho\xi^t = -\frac{{\bar\varepsilon}^2}
  {\kappa_{H}^2}\partial_\varphi{}^2\xi^t
   = -\frac{1}{\kappa_{H}^2}{\bar\partial}_t{}^2\xi^t
   \nonumber\\
&\delta_\xi N^\varphi = 0 && \Rightarrow\ \
   {\hat\xi}^\varphi = \frac{\kappa_{H}^2\rho^2}
   {\bar\varepsilon}q^{\varphi\varphi}\xi^t \nonumber\\
&\delta q_{\rho\varphi}=0 && \Rightarrow\ \
    \rho\partial_\rho{\hat\xi}^\varphi 
    = {\bar\varepsilon}\rho^2q^{\varphi\varphi}\partial_{\varphi}{}^2\xi^t 
    - \omega_2\rho^2\xi^t
    = \frac{{\bar\varepsilon}}{\kappa_{H}^2}
    {\bar\partial}_t{}^2\xi^t - \omega_2\rho^2\xi^t
\end{alignat}
where ${\bar\partial}_t =  \partial_t - N^\varphi\partial_\varphi$ and
${\bar\epsilon} = \Omega_H - {\bar\Omega}$.  
As an additional technical step, we must choose a ``moding'' for the parameters
$\xi^t(\varphi - {\bar\Omega}\,t)$; that is, we write
\begin{equation}
\xi_n^t(\varphi - {\bar\Omega}\,t) = \frac{\gamma}{2{\bar\epsilon}}e^{inu} \qquad
    \hbox{with $u = (\varphi - {\bar\Omega}\,t)/\gamma$} ,
\end{equation}
where it is not completely obvious what value $\gamma$ should 
take.\footnote{One natural choice \citep{Carlip6} is $\gamma={\bar\epsilon}$.  
The modes are then those seen by a corotating observer, with the usual
blue-shifted frequencies $\omega \sim  n/N$, where $N$ is the lapse function.}  
We can then read off the central charge from (\ref{center}), obtaining
\begin{equation}
c = \frac{3A}{2\pi G\kappa_H}\frac{\bar\epsilon}{\gamma} .
\label{cc}
\end{equation}

To use the canonical Cardy formula (\ref{canon}), we must be a bit careful of the
meaning of the temperature $T$.  Again following \citep{Guica}, we note that 
fields in the Frolov-Thorne vacuum have a coordinate dependence
\begin{equation}
\Phi \sim e^{im\varphi - i\omega t}= e^{in_Lu - in_Rt} ,
\end{equation}
where the ``left'' and ``right'' occupation numbers are $n_L = \gamma m$, 
$n_R = \omega - n_L{\bar\Omega}/\gamma$.  The Boltzmann factor is thus
\begin{equation}
e^{-\beta(\omega - m\Omega_H)} = e^{-\beta n_R - ({\bar\epsilon}\beta/\gamma) n_L} .
\end{equation}
From this, we can read off the appropriate CFT temperature for the ``left'' $u$ modes,
\begin{equation}
T_L = \frac{\gamma}{{\bar\epsilon}\beta} = \frac{\gamma}{\bar\epsilon}T_H
        = \frac{\gamma}{\bar\epsilon}\frac{\kappa_H}{2\pi} ,
\label{temp}
\end{equation}
where $T_H$ is the Hawking temperature (\ref{BH}).  Inserting (\ref{cc}) and (\ref{temp})
into the Cardy formula (\ref{canon}) and restoring factors of $\hbar$, we obtain
\begin{equation}
S_{\scriptscriptstyle\mathit{BH}} = \frac{A}{4\hbar G} ,
\end{equation}
the correct Bekenstein-Hawking entropy.

\section{What are the states?}

Recall that one of the original motivations for studying black hole thermodynamics was
to learn something about quantum gravity.  These results suggest, though, that the 
counting of states is universal, depending only on a few characteristics of the symmetry
of the horizon.  Contrary to our hopes, black hole statistical mechanics may not, in
the end, tell us too much about how to quantize general relativity.

One way to understand this conclusion is the following.  In the usual Dirac
treatment of constraints, we would demand that physical states be annihilated
by the constraints,
\begin{equation}
{\hat H}[\xi^\perp,{\hat\xi}^i] |\mathit{phys}\rangle = 0 .
\end{equation}
But if the algebra of constraints contains a Virasoro subalgebra with a nonvanishing
central charge $c$, this condition is inconsistent with the Virasoro algebra (\ref{Vir}).
We know how to handle this problem in conformal field theory, of course \citep{CFT}:
for example, we can demand that only the positive frequency piece of the constraints 
annihilate physical states.  The upshot, though, is that, in the language of 
\citep{Carlip8}, a number of ``would-be pure gauge'' degrees of freedom are no
longer eliminated, but rather become dynamical at the horizon.

A somewhat analogous---although not exactly equivalent---situation occurs in
ordinary field theory.  When a global symmetry is spontaneously broken, a collection 
of massless Nambu-Goldstone bosons appears, one for each ``broken'' symmetry 
generator.  It is well known that these degrees of freedom can be viewed as 
excitations along the ``would-be symmetries'' that are broken by the vacuum.  
But while Goldstone's theorem gives us a good deal of information about these 
bosons, it does not tell us how they are built from the elementary quanta 
of the theory.  The analogy with black hole states would be strengthened if one 
could explicitly derive the Cardy formula as a measure for the ``broken'' 
degrees of freedom; for now, I do not know how to do so.

Let me close with a final question: how might we tell whether the approach described 
here is, in fact, correct?  I certainly don't have a complete answer, but there are
several avenues of research that might help:
\begin{itemize}
\item If a CFT dual explains the universality of black hole entropy, then the conformal
symmetry must be present, if perhaps disguised, in other more ``microscopic'' 
approaches to black hole statistical mechanics---those from string theory and 
loop quantum gravity, for instance.  In \citep{Carlip7}, some evidence is given for 
a connection to the BTZ black hole, and through that to certain string theory approaches,
but much more remains to be done.
\item If black hole degrees of freedom have a Goldstone-like description, then it
should be possible to obtain an effective field theory for the boundary degrees of
freedom directly from general relativity.  This can be done in three spacetime
dimensions \citep{Carlip9}; whether this approach can be extended to higher dimensions
is uncertain.
\item It ought to be possible to couple boundary degrees of freedom to matter and 
connect the results to Hawking radiation.  Again, there are some interesting results 
\citep{Emparan2}, but so far only in three dimensions.
\item The central charge (\ref{cc}) depends on the horizon area, so the model 
described here cannot in itself describe processes such as black hole evaporation,
in which this area changes.  One speculative idea \citep{Carlip2} is that such
processes could be described as flows between conformal field theories.  The direction
of the flow during black hole evaporation is consistent with Zamolodchikov's $c$ 
theorem \citep{Zam}, but for now I can say little more than that.
\end{itemize}


\begin{theacknowledgments}
Portions of this project were carried out at the Peyresq 15 Physics Conference 
with the support of OLAM Association pour la Recherche Fundamentale, 
Bruxelles.  This work was supported by the U.S.\ Department of Energy under grant
DE-FG02-91ER40674.
\end{theacknowledgments}

\end{document}